\documentclass[12pt]{spieman}  
\usepackage{amsmath,amsfonts,amssymb}
\usepackage{graphicx}
\usepackage{setspace}
\usepackage{tocloft}
\usepackage{lineno}
\usepackage{longtable}
\usepackage{caption}
\usepackage{makecell}
\usepackage{algorithm}
\usepackage{algorithmic}
\usepackage{threeparttablex}

\title{Segmentation variability and radiomics stability for predicting  Triple-Negative Breast Cancer subtype using Magnetic Resonance Imaging}

\author[a, b, *]{Isabella Cama}
\author[c]{Alejandro Guzmán}
\author[a, d]{Cristina Campi}
\author[a, d]{Michele Piana}
\author[c, e]{Karim Lekadir}
\author[a]{Sara Garbarino}
\author[c, f]{Oliver Díaz}
\vspace{-0.25cm}

\affil[a]{Universit\`a di Genova, Dipartimento di Matematica, via Dodecaneso 35, Genova, Italy, 16146}

\affil[b]{Università di Genova, Dipartimento di Neuroscienze, Riabilitazione, Oftalmologia, Genetica e Scienze Materno-Infantili, Largo Paolo Daneo 3, Genova, Italy, 16132}

\affil[c]{Universitat de Barcelona, Departament de Matemàtiques i Informàtica, Gran Via de les Corts Catalanes, 585, Barcelona, Spain, 08007}

\affil[d]{IRCCS Ospedale Policlinico San Martino, Largo Rosanna Benzi 10, Genova, Italy, 16132}

\affil[e]{Institució Catalana de Recerca i Estudis Avançats (ICREA), Passeig de Lluís Companys, 23, Barcelona, Spain, 08010}

\affil[f]{Universidad Autónoma de Barcelona, Computer Vision Center, Plaça Cívica, Bellaterra (Cerdanyola del Vallès), Spain, 08193}

\cftpagenumbersoff{figure}
\cftpagenumbersoff{table} 
\begin{document} 
\maketitle

\begin{abstract}
Most papers caution against using predictive models for disease stratification based on unselected radiomic features, as these features are affected by contouring variability.  Instead, they advocate for the use of the Intraclass Correlation Coefficient (ICC) as a measure of stability for feature selection. However, the direct effect of segmentation variability on the predictive models is rarely studied.
This study investigates the impact of segmentation variability on feature stability and predictive performance in radiomics-based prediction of Triple-Negative Breast Cancer (TNBC) subtype using Magnetic Resonance Imaging. 
A total of 244 images from the Duke dataset were used, with segmentation variability introduced through modifications of manual segmentations. For each mask, explainable radiomic features were selected using the Shapley Additive exPlanations method and used to train logistic regression models. Feature stability across segmentations was assessed via ICC, Pearson’s correlation, and reliability scores quantifying the relationship between feature stability and segmentation variability.
Results indicate that segmentation accuracy does not significantly impact predictive performance. While incorporating peritumoral information may reduce feature reproducibility, it does not diminish feature predictive capability. Moreover, feature selection in predictive models is not inherently tied to feature stability with respect to segmentation, suggesting that an overreliance on ICC or reliability scores for feature selection might exclude valuable predictive features.
\end{abstract}


\noindent \footnotesize\textbf{*} Send correspondence to Isabella Cama, email: \linkable{isabella.cama@edu.unige.it}

\fontsize{12pt}{14pt}\selectfont
\section{Introduction}
\label{sec:introduction} 
Breast cancer is a complex and heterogeneous disease, with multiple molecular subtypes that hamper accurate prediction of disease evolution and the development of targeted treatments~\cite{johnson2021molecular}. 

Triple Negative Breast Cancer (TNBC) is defined by the absence of estrogen receptor (ER), progesterone receptor (PR), and HER2 overexpression, accounting for approximately 15–20\% of all breast cancers and disproportionately affecting younger and African American women~\cite{foulkes2010triple,bianchini2016triple}. It is known to present a higher grade, earlier recurrence, and worse overall prognosis compared to other subtypes~\cite{foulkes2010triple}, thus representing the most challenging breast cancer subtype to treat. 

Subtype determination has always relied on methods such as immunohistochemistry, staining, and fluorescence in situ hybridization \cite{johnson2021molecular}. 
Recent advances in literature indicate the potential for predicting molecular subtypes using image-based features through machine learning (ML) techniques \cite{sha2022mri}. These features, known as radiomic features, number in the thousands and can capture detailed information about morphology, intensity, and texture within a specific region of interest (ROI), such as a tumor. The standard radiomics workflow requires an initial step of image segmentation~\cite{gillies2016radiomics,zwanenburg2020image}, often relying on the annotator’s skill and the clarity of ROI boundaries~\cite{rizzo2018radiomics}. Then, radiomic features are extracted from the segmented images and irrelevant or redundant features are discarded through feature selection procedures~\cite{conti2021radiomics}. Finally, predictive models are trained, and their application to diagnosis, prognosis, and treatment response prediction has been increasingly reported in literature of breast cancer \cite{TAGLIAFICO202074}.

Son et al.~\cite{son2020prediction} conducted a study on synthetic mammography, reconstructed from digital breast tomosynthesis, to predict the molecular subtype using both clinical and radiomic features. An elastic-net logistic regression \cite{elasticnet} model was trained to create the radiomics signature of each lesion. The results show that the combination of radiomics and clinical data outperformed the prediction using only clinical data, suggesting that radiomic signatures could serve as biomarkers for TNBC. In the case of Magnetic Resonance Imaging (MRI), Leithner et al.~\cite{leithner2020non} combined radiomic data extracted from dynamic contrast-enhanced (DCE) MRI and apparent diffusion coefficient (ADC) map to differentiate TNBC from other subtypes. Overall, these studies showed that the radiomic approach could support the identification of TNBC patients, hence contribute to the improvement of treatment planning by offering a non-invasive way of analyzing tumor biology. 

However, the widespread adoption of radiomics in clinics is hampered by issues related to features stability \cite{cama2024study, cama2023segmentation}. Factors that could affect radiomic features computation can be found in the image acquisition and reconstruction phase, in the image pre-processing steps, and in the segmentation of the region of interest from which radiomic features are extracted. Scaco et al. \cite{Scalco_2022} reported a list of papers that evaluated the effects of segmentation on radiomics stability, with just one of them considering applications to breast cancer imaging.
Granzier et al. \cite{granzier} studied the robustness of radiomic features, extracted by two software tools, with respect to variability in manual segmentation of breast tumor on MRI. Although a threshold value of 0.90 for the Intraclass Correlation Coefficient (ICC) \cite{Koo2016} was chosen to determine feature robustness, its significance for radiomic models in predicting patient outcomes was not investigated.
Other studies addressed the problem of assessing the robustness of radiomic features by segmentation perturbation \cite{Zwanenburg, Yang, Tixier} for various applications, various type of imaging (Computed Tomography (CT), Positron Emission Tomography (PET), and MRI), and for different diseases (lung cancer, head and neck squamous cell carcinoma, glioblastoma).  All these studies call for caution in the use of predictive models involving radiomic features implicated with contouring variability within the context of disease stratification and risk assessment, although no prediction experiment is reported.
Kothari et al. \cite{Kothari} highlighted that the selection of robust features from masks delineated by different clinicians allows survival models to retain their prognostic ability.
Among the papers addressing segmentation variability in a more systematic way, along with predictive tasks, Poirot et al. \cite{Poirot} studied how differences among segmentations affect radiomic features in neuroimaging, in a cohort of T1-weighted and diffusion tensor images of sleep-deprived patients. The robustness and reproducibility of radiomic features were assessed using ICC for descriptive purposes, while the performance of the predictive model is only expressed in terms of accuracy, and no statistical analysis is reported to evaluate feature robustness in relation to predictive performance. 
Liu et al. \cite{LIU202011} showed that for CT images of  oropharyngeal cancer, radiomic features varied a lot when the ROIs were not well segmented (ICC). No statistical test is reported for the significance of the results, and no quantification of the segmentation variability is presented. Moreover, the authors do not examine the behavior of the same group of features across variations in masks, nor do they attempt to generate a new prognostic signature; instead, they train univariate models for prediction.

In summary, the existing literature uses the ICC to evaluate radiomic features' reproducibility, as a synonym of stability, reliability and robustness, across different segmentation masks, and to perform feature selection. Only in few cases a prediction task is performed on the selected robust features, usually on a fixed data split. However, the ICC-based approach does not consider the possibility of significant quantitative differences in the various segmentation outcomes, implicitly assuming that the segmentation agreement is always high. 
Moreover, when analyzing feature stability exclusively with respect to segmentation, i.e., outside the context of prediction, specific features may exhibit a lack of robustness, which would be mostly mitigated during prediction, due to the feature scaling required by ML methods.

In this study, we investigate how feature stability and prediction performance are impacted by segmentation variability, focusing on the predictive performance of radiomics-based ML models designed to differentiate TNBC from other molecular subtypes based on MRI-derived features. 
The stability of the feature selected by these models is reported via ICC, for comparison with existing literature, Pearson's correlation, and through the method we described in \cite{cama2023segmentation}, which introduces four quantitative scores measuring feature stability in terms of consistency, robustness, instability, and quality of feature computation and explicitly accounting for possible variability between results of the segmentation process.
We focused on identifying radiomic features that are consistently predictive of TNBC across a population of breast cancer patients, and we explored the prediction performances and the stability of the most significant radiomic features across different segmentation masks. 

While this investigation focuses on a specific application, its findings could potentially elucidate the general behavior of radiomic features, providing a broader understanding of the principles underlying radiomics-based analyses.

\section{Materials and Methods}\label{sec:methods}
\subsection{Data collection}
Consent or waiver for data usage was not required since all data were obtained in de-identified form from the publicly available Duke-Breast-Cancer-MRI dataset~\cite{duke_dataset}, hosted on The Cancer Imaging Archive \cite{tcia}. For each patient, the Duke-Breast-Cancer-MRI dataset contains DCE-MR images from multiple time points, capturing both pre-contrast and post-contrast phases. Imaging data were collected using various scanner models from GE MEDICAL SYSTEMS, MPTronic, and SIEMENS, including the Avanto, Optima MR450w, SIGNA EXCITE, SIGNA HDx, SIGNA HDxt, Skyra, Trio, and TrioTim. The dataset also contains demographic, clinical, pathology and  treatment information, and outcomes of patients (e.g., response to treatment, recurrence, follow-up).
Pre-operative DCE-MR first-post-contrast images were used for this study. The patients involved correspond to the 251 selected by Caballo et al. \cite{caballo2023four}. Seven patients were excluded due to issues related to DICOM metadata. Therefore, a cohort of 244 patients was analyzed for this study, including 71 with TNBC and 173 non-TNBC (30\% vs. 70\% of the dataset, respectively). The dataset was preliminary split into a 70\%-30\% stratified proportion for training and testing of the ML models, for 130 random data splits.

\subsection{Segmentation}\label{sec:segmentation}
\begin{figure}[h!]
    \centering
    \begin{tabular}{ccc}
        \includegraphics[width=0.3\linewidth]{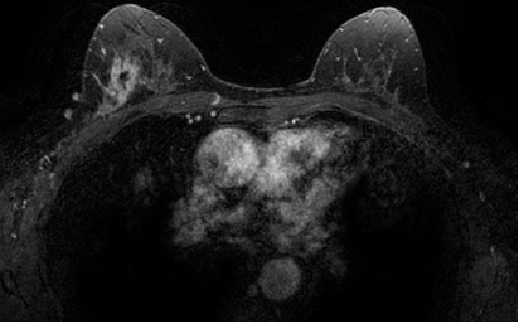} &
        \includegraphics[width=0.3\linewidth]{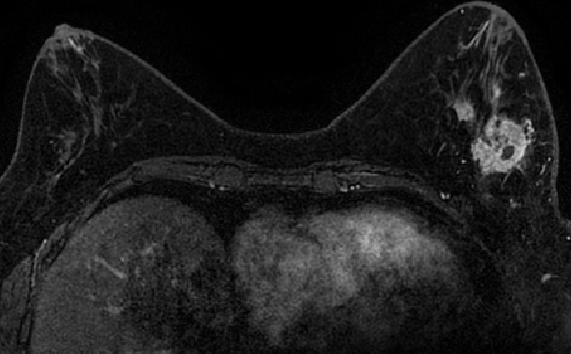} &
        \includegraphics[width=0.3\linewidth]{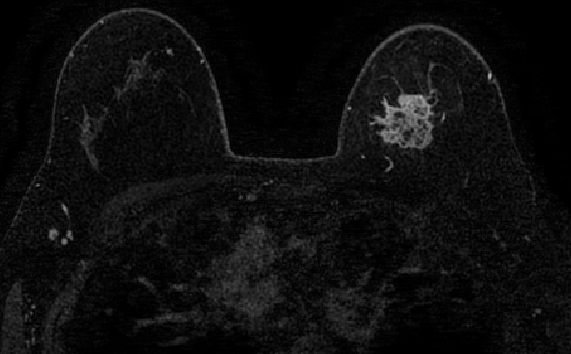} \\
        \includegraphics[width=0.3\linewidth]{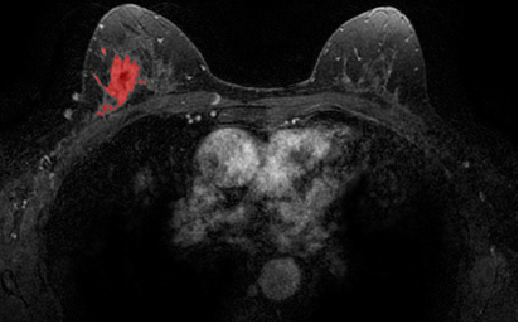} &
        \includegraphics[width=0.3\linewidth]{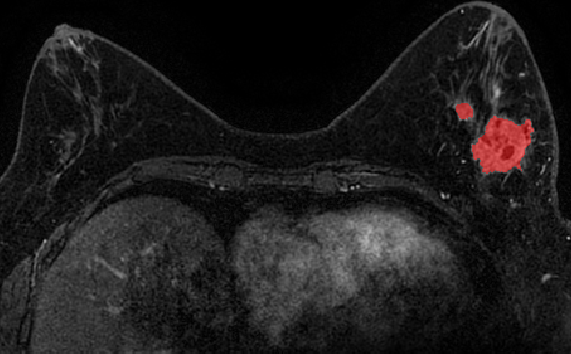} &
        \includegraphics[width=0.3\linewidth]{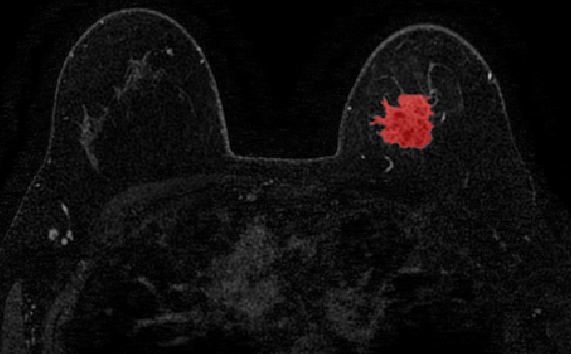} \\
        \includegraphics[width=0.3\linewidth]{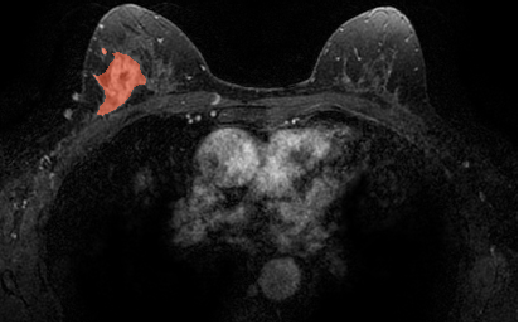} &
        \includegraphics[width=0.31\linewidth]{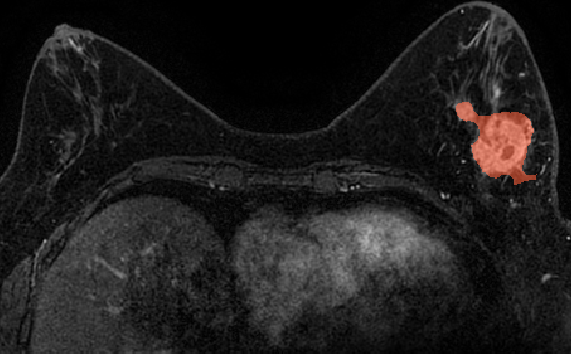} &
        \includegraphics[width=0.3\linewidth]{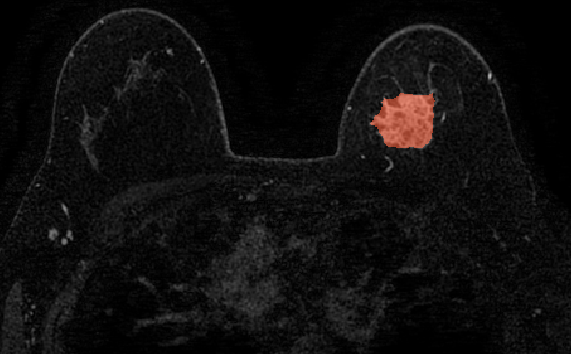} \\
        \includegraphics[width=0.3\linewidth]{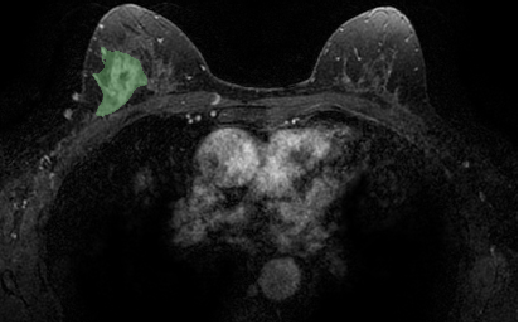} &
        \includegraphics[width=0.3\linewidth]{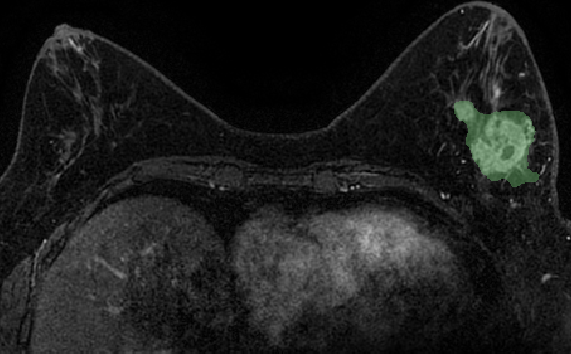} &
        \includegraphics[width=0.3\linewidth]{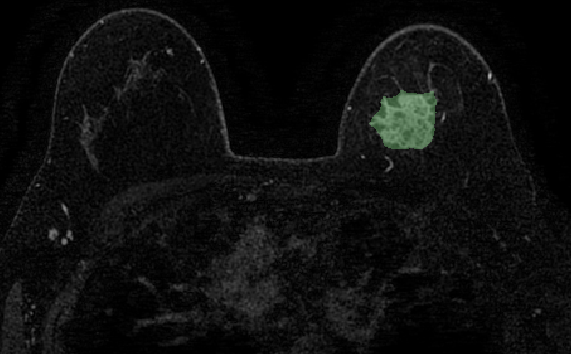} \\
        \includegraphics[width=0.3\linewidth]{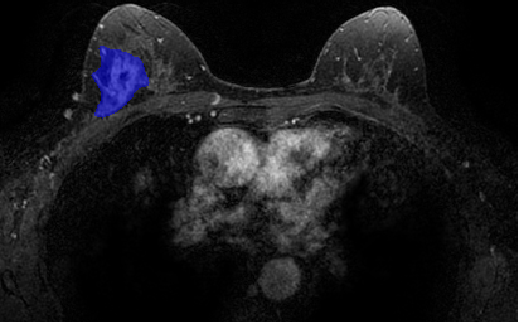} &
        \includegraphics[width=0.3\linewidth]{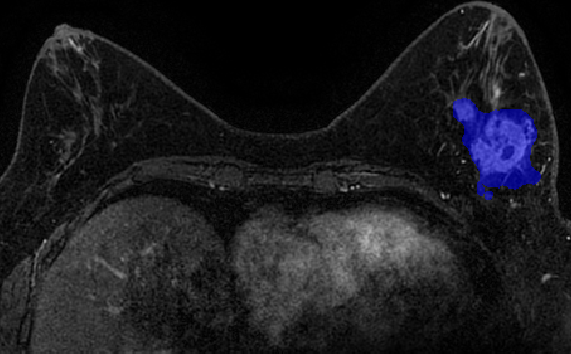} &
        \includegraphics[width=0.3\linewidth]{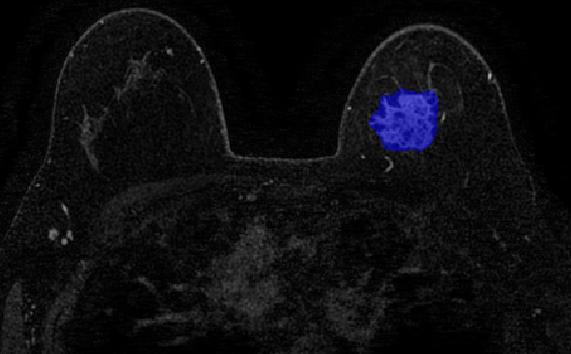} \\
        \includegraphics[width=0.3\linewidth]{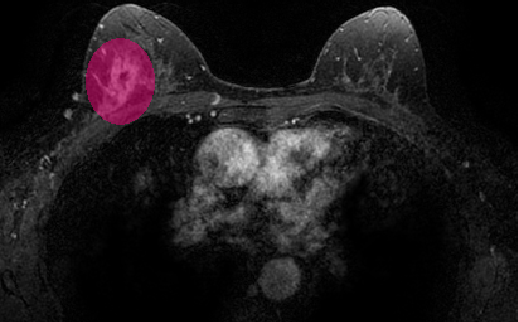} &
        \includegraphics[width=0.3\linewidth]{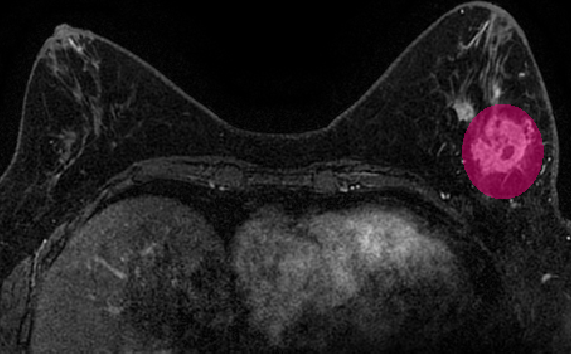} &
        \includegraphics[width=0.3\linewidth]{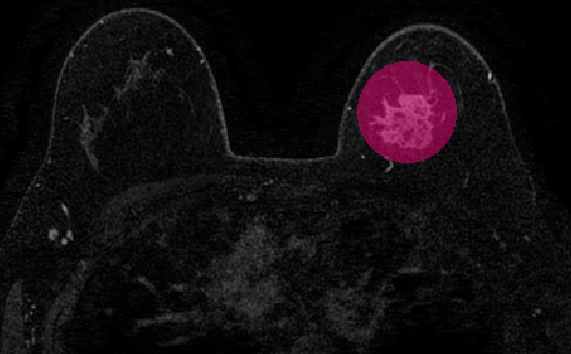} \\
    \end{tabular}
    \caption{Left to right: MR slices of three different patients (first-post-contrast image). Top to bottom: original image, manual tumor segmentation (red), closing 08 mask (orange), closing 07 (green), closing 06 (blue), and ellipsoid 04 (magenta). For the definition of `closing mask' see Section \ref{sec:segmentation}.}
    \label{fig:overall}
\end{figure}

Manual segmentation of MR images was provided by Caballo et al~\cite{caballo2023four}. To assess the stability of the features with respect to segmentation accuracy, we introduced variability in the manual segmentation mask by simulating four other annotations per case. Our aim was to obtain an average mean DSC across all segmentations ranging from 0.4 to 0.8 with respect to the original mask. This was achieved via the morphological operation of closing. Using the \texttt{ball} structuring element from the \texttt{skimage.morphology} library \cite{van2014scikit} with varying kernel sizes, we gradually enlarged the manual segmentation mask to achieve various mean Dice Similarity Coefficients (DSC), including more and more portions of the peritumoral region. Specifically, we applied the following operations:
\begin{itemize}
    \item dilation and erosion operations, using a kernel size of 5, to obtain mean DSC 0.8;
    \item dilation and erosion, using a kernel size of 9, to obtain mean DSC 0.7;
    \item  dilation with a kernel size of 11 and erosion with a kernel size of 9, to obtain mean DSC 0.6;
    \item finally, the manual segmentation was replaced by the ellipsoid contained in the ROI box provided by the dataset along with the images, to obtain mean DSC 0.4.
\end{itemize}
The implemented modifications of the manual masks allowed for consideration of a broad range of potential contour variations, as the segmentation obtained through morphological operations exhibited significant variability. For instance:
  \begin{itemize}
  	\item certain tumors were originally located near the skin, causing the modified mask to extend beyond the patient's body;
  	\item segmentation of tumors with multiple lesions resulted to be grouped into a single mask (Figure \ref{fig:overall}, central column);
  	\item spike-like structures were enhanced (Figure \ref{fig:overall}, in orange).
  \end{itemize}
   Additionally, the ellipsoid mask served as a geometric approximation of tumor segmentation (DSC $0.4$), fulfilling the same function as a ROI box by providing a highly simplified approximation, while offering a more natural shape, closer to a potential manual segmentation. 
   
   In the rest of the document the manual mask and its modifications are referred to as manual, closing $08$, closing $07$, closing $06$, and ellipsoid $04$, respectively (see Figure~\ref{fig:overall}), referring to the mean DSC value reached by each morphological operation applied to the original mask.

\subsection{Radiomic feature extraction}
Radiomic features were extracted from the segmentation masks after image z-score normalization and fixed-bin count discretization with 50 bins using PyRadiomics 3.1.0 library~\cite{pyradiomics}.
We extracted 1130 radiomic features including shape, texture, matrix-based, wavelet, and Laplacian of Gaussian (LoG) features ($\sigma = $ 1, 2, 3). 

Shape-based features describe the geometric properties of a region of interest, such as volume, surface area, and compactness. First-order statistics capture the distribution of voxel intensities within the image, summarizing global intensity characteristics. Gray Level Co-occurrence Matrix (GLCM) features assess spatial relationships between intensity values, reflecting textural uniformity and contrast. Gray Level Run Length Matrix (GLRLM) features quantify the length and distribution of consecutive voxel intensity runs, characterizing texture smoothness. Gray Level Size Zone Matrix (GLSZM) features evaluate the size and distribution of homogeneous intensity zones. Neighbouring Gray Tone Difference Matrix (NGTDM) features measure texture strength and contrast based on intensity differences with neighboring voxels. Gray Level Dependence Matrix (GLDM) features capture the degree to which voxels depend on neighboring intensities, quantifying image granularity and texture complexity. 
 
All these features were extracted from the original and filtered images. Wavelet-HHH, wavelet-HLH, and wavelet-LLL filters are types of wavelet transform filters \cite{wavelet} where HHH applies high-pass filtering across all three dimensions to capture high-frequency features, HLH applies a combination of high and low-pass filtering in different dimensions to extract mixed frequency information, and LLL applies low-pass filtering across all dimensions to capture the overall low-frequency components. On the other hand, LoG \cite{log} is an edge detection method where $\sigma$ values denote the scale of Gaussian smoothing applied before computing the Laplacian to detect edges at varying levels of detail. Extensive documentation of radiomics features can be found in \cite{pyradiomics}.

For the analysis, features were normalized using z-score. Then, they were harmonized to limit the potential bias introduced by the differences in signal-to-noise ratio caused by the manufacturer model. Harmonization was performed using the parametric version of the ComBat method \cite{combat, combatJMI}, which yields a transformation of the feature distributions according to the variable being tested (manufacturer model) using additive and multiplicative batch effects.

\subsection{Clinical features}
  Clinical features were utilized for prediction purposes to enable comparisons with radiomics-based models. Specifically, this work considered demographical variables, i.e., age, menopause at diagnosis, ethnicity, and metastatic state at presentation, and biopsy variables, including tubule formation, nuclear grade, and mitotic rate. These biopsy variables are clinically used for breast cancer grading and are strongly associated with tumor aggressiveness. While they do not directly determine tumor subtype, they may be naturally linked to it, as higher histologic grades often correlate with more aggressive subtypes. For this reason, the model based on biopsy variables could serve as a reference for comparison in subtype prediction.

\subsection{Machine Learning models}
For each data split, the imbalanced data issue (TNBC vs. non-TNBC) was addressed prior to the training process via the synthetic minority oversampling technique \cite{smote} (SMOTE), a widely used method for random oversampling of tabular data, like radiomic features. 
A preliminary selection of the $50$ most informative features was performed through ANOVA F-values computation, used to rank features based on their relevance to the target variable. 
Then, a Logistic Regression model with L$1$-norm penalty \cite{Nick2007} was trained for the classification task TNBC vs. non-TNBC with $5$-fold cross-validation. The regularization parameter $C$, to control feature selection strength, was set to $1$. L$1$-norm penalty is known to promote sparsity by selecting only the most relevant features for the model, thus identifying the radiomic signature of each lesion.

\begin{figure}[t]
\centering
    \includegraphics[width=0.8\linewidth]{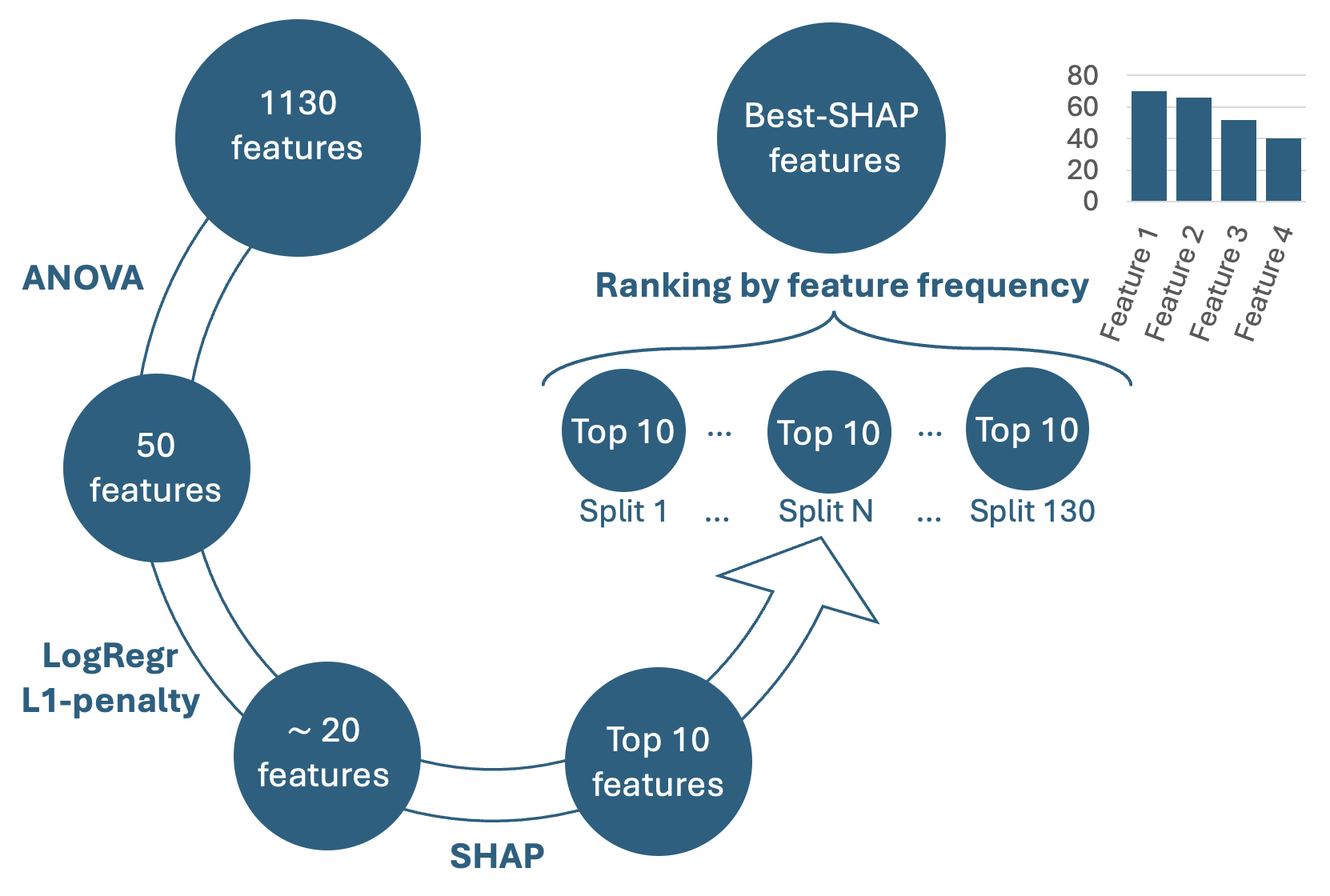}
    \caption{Flowchart of the feature selection methodology employed for this study, based on SHAP explainability algorithm.}
    \label{fig:flowchart_feature_selection}
\end{figure}

After the training process, we used SHAP algorithm (SHapley Additive exPlanations~\cite{lundberg2017unified}) to collect the most explicative features for each model (each trained on a different data split). SHAP is a feature importance tool, based on a game theoretic approach, used in ML for explaining the output of a model by quantifying the importance of each feature. SHAP identifies the most relevant features that contribute to the model's predictions by calculating SHAP values for each feature in the dataset: features with higher SHAP values are considered more influential in the model's predictions, while features with lower SHAP values have less impact.
Specifically, for each trained model, we identified the top $10$ features selected by SHAP. From the aggregated set of top SHAP-selected features across all models, we further selected those with the highest frequency of occurrence, that appeared at least $15$ times overall (the 'best-SHAP features' from now on). These features can be considered collectively predictive, as they contributed to the predictive performance across the whole dataset. A diagram outlining the SHAP feature selection methodology employed in this study is shown in Fig. \ref{fig:flowchart_feature_selection}, while detailed implementation is reported in Algorithm \ref{best_shap_algo}. The best-SHAP features were then used to train Logistic Regression models, one for each data split, for comparison with the baseline models, described at the beginning of this section. This procedure was repeated for each feature set, extracted from the segmentation masks described in Section \ref{sec:segmentation}. In the following, 'baseline-manual', 'baseline-closing $08$', 'baseline-closing $07$', 'baseline-closing $06$', and 'baseline-ellipsoid $04$' will refer to baseline models trained with features extracted from the corresponding mask, while 'best-SHAP-manual', 'best-SHAP-closing $08$', 'best-SHAP-closing $07$', 'best-SHAP-closing $06$', and 'best-SHAP-ellipsoid $04$'  will indicate models trained with best-SHAP features derived from the corresponding baseline model. 
  All ML models were developed in Python (v3.12.2).

  	\begin{algorithm}[h!]
  	\caption{Best-SHAP feature selection \label{best_shap_algo}}
	Input: radiomic features extracted from the manual segmentation\\
  	{\bf{Step $\boldsymbol{1}$: preliminary feature selection and model training}} \\
  	For each data split (training set):
  	\begin{enumerate}
  		\item compute ANOVA F-values to rank features based on their relevance to the target (TNBC)
  		\item select the first $50$ features
  		\item train the predictive model with such features (each LogReg model further selects approximately $20$ features due to L$1$ penalty)
  		\item collect the top $10$ features considered explainable for the current model by SHAP method
  	\end{enumerate}
  	{\bf{Step $\boldsymbol{2}$: best-SHAP feature selection}}  \\
	From the aggregated set of SHAP-selected features identified in Step $1$:
	\begin{enumerate}
		\item count the frequency of occurrence of each feature
		\item rank them in descending order
		\item select the ones that appeared at least 15 times overall: these are the best-SHAP features for the manual segmentation
	\end{enumerate}
	Repeat steps $1$ and $2$ for each feature set (closing $08$, $07$, $06$, ellipsoid $04$).
  \end{algorithm}

\subsection{Statistical analysis}

Univariate statistical analysis was performed to evaluate the potential signiﬁcance of individual descriptors in discriminating TNBC vs. non-TNBC cases. 
Continuous variables were analyzed with two-sided Kolmogorov-Smirnov (KS) test, while discrete variables with the Pearson’s chi-squared test. Bonferroni correction for multiple comparisons was applied.

Prediction models, leveraging clinical, radiomic and the best-SHAP features, underwent rigorous evaluation for various performance metrics: accuracy, recall, specificity, balanced accuracy (which accounts for dataset imbalance by computing the mean of recall and specificity), and Area Under the Receiver Operating Characteristic Curve (ROC-AUC). The performances reported in this manuscript are averaged across data splits and 95\% conﬁdence intervals (CI) are reported. 
Kolmogorov-Smirnov test was used to evaluate the difference in score distributions from different ML models, with Bonferroni correction for multiple comparisons.

To check feature stability/robustness across different segmentation masks, the most relevant features for the models were evaluated via ICC \cite{Koo2016} and Pearson's correlation coefficient. 
The ICC measures agreement between measurements and, in radiomics, is often used to assess feature reproducibility across different segmentation masks (representing the raters). This study utilized a two-way random effects model, generally denoted as ICC($2$), which is appropriate when all raters evaluate all subjects and both raters and subjects are considered random effects. The computation accounts for variability between subjects and raters, as well as error variance, which may result from differences in rater performance or other unaccounted noise. 
ICC and Pearson's correlation were computed pairwise using the features extracted from the manual mask and each of its modifications (DSC $08$, $07$, $06$, $04$).

To evaluate feature stability, we also used the scores proposed in \cite{cama2023segmentation}, measuring the numerical relationship between feature stability and tumor segmentation. These 'reliability' scores are based on a quantitative assessment of segmentation variability and the relative error on feature values, where 'reliability' refers to feature stability with respect to segmentation variability:

\begin{itemize}
    \item the quality score indicates when segmentation agreement and feature computation accuracy are simultaneously high;
    \item the consistency score highlights anti-linear correlation of feature error and segmentation result (the error grows while the DSC decreases);
    \item the robustness score indicates independence of feature value from segmentation variability;
    \item the instability score shows high dependence of feature value on minor segmentation variations.
\end{itemize} 

  These scores can serve as quantitative parameters for a reliability/stability assessment process in radiomics. Specifically, for each feature, the scores provide measures of the collective behavior of that feature on a group of patients, by quantifying the proportion of patients with specific trend of feature stability. However, their interpretation is tied to the distribution of the DSC along the x-axis. In this study, the DSC was deliberately adjusted to achieve specific mean values, concentrating the analyses on specific portions of the x-axis for each case.

\section{Results}\label{sec:results}
Table \ref{tab:segmentation_dice} displays the means and standard deviations of the DSC obtained for the various modifications of the manual segmentation mask, computed across the patients involved in the study.

\begin{table}[ht]
\centering
\caption{Mean $\pm$ standard deviation of the DSC obtained for each modification of the manual segmentation mask, computed across all the patients.}
\label{tab:segmentation_dice}
\begin{tabular}{|lc|}
\hline
\textbf{Segmentation mask} & \textbf{Mean DSC $\pm$ standard deviation} \\
\hline
Closing 08 & 0.81 ± 0.13 \\
\hline
Closing 07 & 0.69 ± 0.12 \\
\hline
Closing 06 & 0.60 ± 0.12 \\
\hline
Ellipsoid 04 & 0.40 ± 0.13 \\
\hline
\end{tabular}
\end{table}
 \vspace{0.3cm}
The univariate feature analysis, performed on the whole dataset, showed that no single feature independently exhibited a significant difference between the TNBC and non-TNBC groups based on the Kolmogorov-Smirnov and Chi-squared tests (p-value$>$ 0.05).

\begin{figure}[h!]
    \centering
        \includegraphics[width=\linewidth]{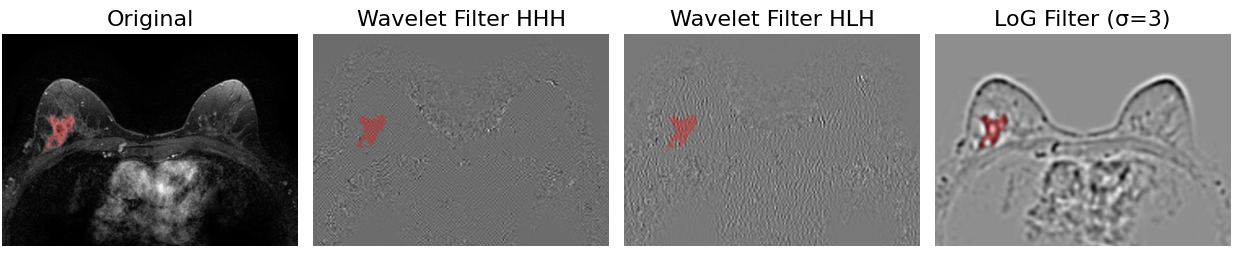}\\
        \includegraphics[width=\linewidth]{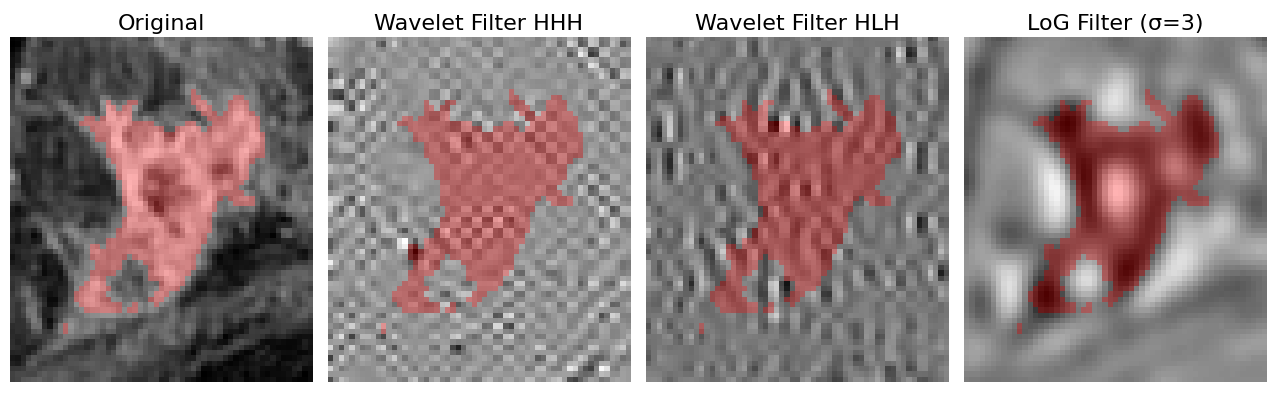}
        
        \caption{Top row, from left to right: example slice of original DCE-MR image, Wavelet-HHH filtered image, Wavelet-HLH filtered image, and LoG $\sigma=3$ filtered image. Bottom row, from left to right: zoomed view of the manual segmentation mask on the original and filtered images. Tumor segmentation is shown in red.}
        \label{fig:wavelet}
    \end{figure}

Table \ref{tab:commonShap} lists the best-SHAP-features for each model, identified as the most overall explicative features for each segmentation mask, as described in Section \ref{sec:methods}. Only four of these features were commonly selected by the models and are reported in bold in Table \ref{tab:commonShap}: High-Gray-Level-Zone Emphasis (from LoG $\sigma=3$ image, GLSZM matrix), Large-Dependence High-Gray-Level Emphasis (from original image and GLDM matrix), Zone Entropy (from wavelet-HHH image and GLSZM matrix), and Skewness (from wavelet-HLH image, first order features). Figure \ref{fig:wavelet} reports an example of wavelet-HHH, wavelet-HLH, and LoG $\sigma=3$ filtered images, for a visual reference of the filtering effect.

\renewcommand{\arraystretch}{2} 
\setlength{\tabcolsep}{2.5pt} 
{\fontsize{8.2}{8.2}\selectfont
\begin{longtable}{|lllccccc|}
\caption{List of the best-SHAP features identified as explicative through SHAP method for each model. 'X' indicates the model for which the feature has been selected as explicative. Features that are selected by all the models are highlighted in bold.}
\label{tab:commonShap} \\
\hline
\textbf{Image Filter} & \textbf{Matrix} & \textbf{Feature Name} & \textbf{Manual} & \textbf{Closing 08} & \textbf{Closing 07} & \textbf{Closing 06}  & \textbf{Ellipsoid 04} \\
\hline
\endfirsthead
\hline
\textbf{Image Filter} & \textbf{Matrix} & \textbf{Feature Name} & \textbf{Manual} & \textbf{Closing 08} & \textbf{Closing 07} & \textbf{Closing 06}  & \textbf{Ellipsoid 04} \\
\hline
\endhead
\hline \multicolumn{8}{r}{\textit{Table \ref{tab:commonShap} continued on next page}} \\ 
\endfoot
\hline
\endlastfoot
LoG $\sigma=2$ & glcm & Cluster Prominence & X & - & - & - & - \\ \hline
LoG $\sigma=3$ & glrlm & Run Entropy & X & X & - & - & - \\ \hline
\textbf{LoG $\sigma=3$} & \textbf{glszm} & \textbf{High-Gray-Level-Zone Emphasis}  & \textbf{X} & \textbf{X} & \textbf{X} & \textbf{X} & \textbf{X} \\ \hline
LoG $\sigma=3$ & glszm & Low-Gray-Level-Zone Emphasis & - & - & - & X & - \\ \hline
LoG $\sigma=3$ & glszm & Small-Area-High-Gray-Level Emphasis & X & X & - & - & - \\ \hline
original & firstorder & Median & - & - & X & - & - \\ \hline
original & firstorder & Minimum & - & X & X & X & X \\ \hline
\textbf{original} & \textbf{gldm} & \textbf{Large-Dependence-High-Gray-Level Emphasis} & \textbf{X} & \textbf{X} & \textbf{X} & \textbf{X} & \textbf{X} \\ \hline
wavelet-HHH & firstorder & Median & - & - & - & - & X \\ \hline
wavelet-HHH & firstorder & Uniformity & - & - & X & - & - \\ \hline
wavelet-HHH & glcm & Cluster Tendency & - & - & - & X & - \\ \hline
wavelet-HHH & glcm & Inverse Variance & X & - & - & - & - \\ \hline
wavelet-HHH & glcm & Joint Energy & - & X & - & - & - \\ \hline
wavelet-HHH & glcm & MCC & X & - & X & - & - \\ \hline
wavelet-HHH & glcm & Maximum Probability & X & X & - & - & - \\ \hline
wavelet-HHH & glcm & Sum Entropy & X & X & X & X & - \\ \hline
wavelet-HHH & gldm & Dependence Entropy & X & X & - & X & X \\ \hline
wavelet-HHH & glszm & Gray-Level Non-Uniformity Normalized & X & - & - & - & X \\ \hline
wavelet-HHH & glszm & Gray-Level Variance & - & - & X & X & - \\ \hline
\textbf{wavelet-HHH} & \textbf{glszm} & \textbf{Zone Entropy} & \textbf{X} & \textbf{X} & \textbf{X} & \textbf{X} & \textbf{X} \\ \hline
wavelet-HHH & ngtdm & Complexity & X & - & X & - & X \\ \hline
wavelet-HHL & firstorder & Skewness & - & - & X & - & - \\ \hline
wavelet-HHL & glcm & MCC & - & X & X & - & X \\ \hline
\textbf{wavelet-HLH} & \textbf{firstorder} & \textbf{Skewness} & \textbf{X} & \textbf{X} & \textbf{X} & \textbf{X} & \textbf{X} \\ \hline
wavelet-HLH & glcm & Cluster Shade & - & - & X & - & - \\ \hline
wavelet-HLH & gldm & Dependence Entropy & X & - & - & - & X \\ \hline
wavelet-HLH & glrlm & Long-Run-High-Gray-Level Emphasis & - & X & X & X & - \\ \hline
wavelet-HLH & glrlm & Short-Run-High-Gray-Level Emphasis & - & X & - & - & - \\ \hline
wavelet-HLH & glszm & Low-Gray-Level-Zone Emphasis & - & - & X & X & X \\ \hline
wavelet-HLH & glszm & Small-Area-High-Gray-Level Emphasis & - & - & X & - & - \\ \hline
wavelet-HLH & glszm & Small-Area-Low-Gray-Level Emphasis & - & - & - & X & - \\ \hline
wavelet-HLH & glszm & Zone Entropy & - & - & - & - & X \\ \hline
wavelet-HLL & glszm & Small-Area-High-Gray-Level Emphasis & - & - & - & - & X \\ \hline
wavelet-LHH & firstorder & Median & - & X & - & - & - \\ \hline
wavelet-LHH & glcm & Autocorrelation & - & X & - & - & - \\ \hline
wavelet-LHH & glszm & Zone Entropy & X & X & X & - & X \\ \hline
wavelet-LHL & firstorder & Interquartile Range & - & - & - & X & - \\ \hline
wavelet-LHL & firstorder & Mean & - & X & X & X & - \\ \hline
wavelet-LHL & firstorder & Median & - & - & - & X & - \\ \hline
wavelet-LHL & firstorder & Skewness & - & X & - & - & - \\ \hline
wavelet-LHL & glcm & Correlation & - & X & X & X & X \\ \hline
wavelet-LHL & glrlm & Long-Run-Low-Gray-Level Emphasis & - & - & - & X & - \\ \hline
wavelet-LLH & glcm & Imc1 & - & - & - & X & - \\ \hline
wavelet-LLH & glcm & MCC & X & X & X & - & - \\ \hline
wavelet-LLL & firstorder & 10th Percentile & X & - & - & - & - \\ \hline
wavelet-LLL & firstorder & Median & - & - & - & X & - \\ \hline
wavelet-LLL & glcm & Cluster Shade & - & X & X & - & - \\ \hline
wavelet-LLL & glcm & MCC & - & - & - & - & X \\ \hline
wavelet-LLL & gldm & Large-Dependence-High-Gray-Level Emphasis & X & - & - & - & X \\ \hline
wavelet-LLL & glrlm & Long-Run-High-Gray-Level Emphasis & - & X & - & - & - \\ \hline
wavelet-LLL & glszm & Small-Area-Low-Gray-Level Emphasis & - & - & - & X & - \\ \hline

\end{longtable}

}

\begin{figure}[h!]
	\centering
    \includegraphics[width=\linewidth]{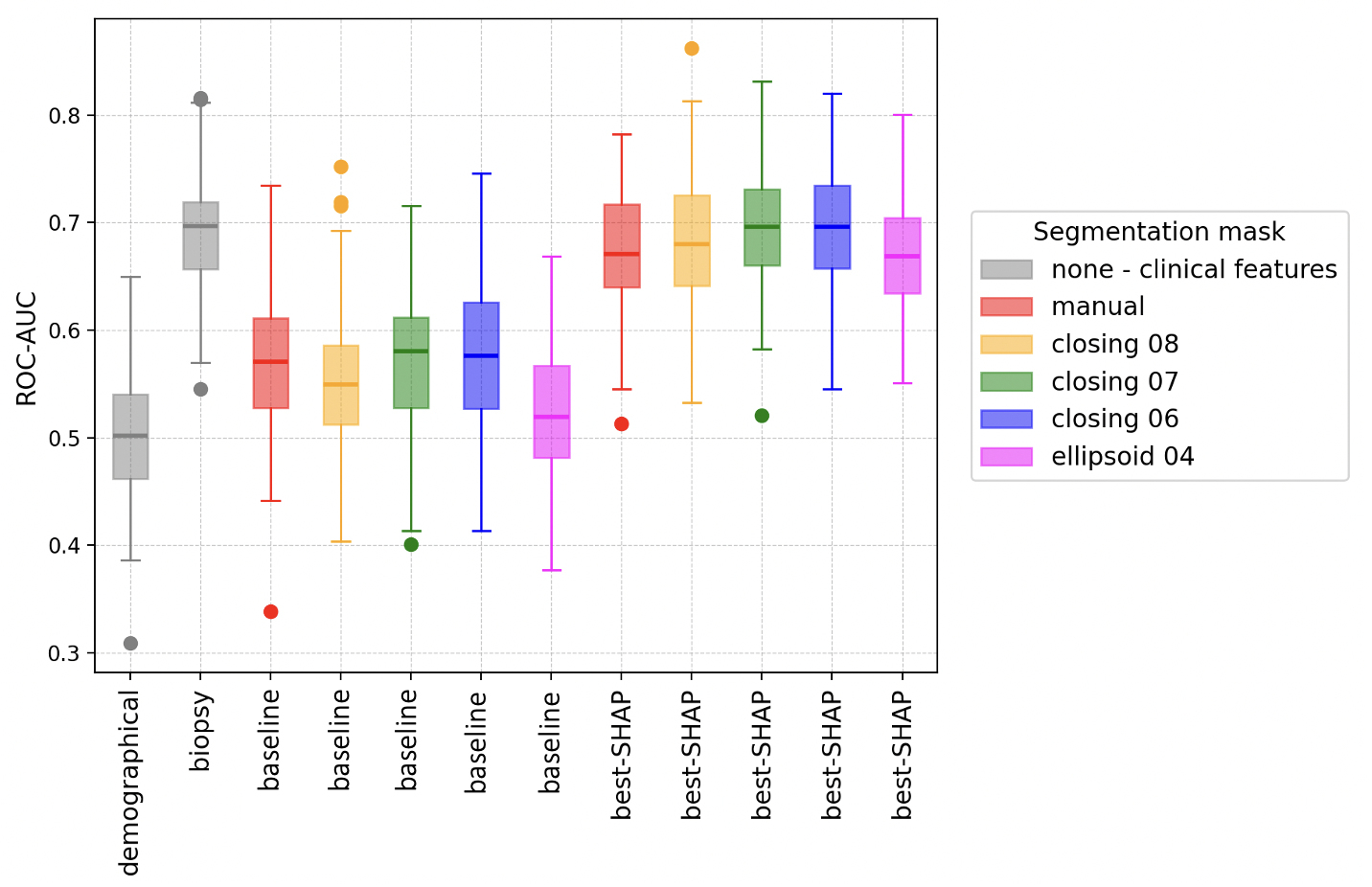}
    \caption{ROC-AUC scores obtained by testing demographical model (boxplot $1$), biopsy model (boxplot $2$), baseline models (boxplots $3$-$7$), and best-SHAP models (boxplots $8$-$12$).}
    \label{fig:roc_clinical_all_shap}
\end{figure}

Figure \ref{fig:roc_clinical_all_shap} illustrates the performance of the prediction models using demographical variables (boxplot $1$) and biopsy variables (boxplot $2$), as well as baseline models (boxplots $3$-$7$) and best-SHAP models (boxplots $8$-$12$). Performance is shown in terms of ROC-AUC. Table \ref{tab:skill_scores} reports detailed statistics, including skill-score means and confidence intervals obtained for each ML model. The prediction using solely demographical variables is random.

\renewcommand{\arraystretch}{2} 

{\fontsize{8.6}{8.6}\selectfont

\begin{longtable}{|lccccc|}

\caption{Performances of demographical model, biopsy model, baseline models, and best-SHAP models for TNBC prediction. Mean skill-scores and their $95$\% confidence interval (within brackets) are reported. In bold, the best mean result for each skill-score.}
\label{tab:skill_scores} \\
\hline
\textbf{Experiment} & \textbf{Accuracy} & \textbf{Balanced Accuracy} & \textbf{Recall} & \textbf{Specificity} & \textbf{ROC-AUC} \\ \hline
\endfirsthead
\hline
\textbf{Experiment} & \textbf{Accuracy} & \textbf{Balanced Accuracy} & \textbf{Recall} & \textbf{Specificity} & \textbf{ROC-AUC} \\ \hline
\endhead
\hline \multicolumn{6}{r}{\textit{Table \ref{tab:skill_scores} continued on next page}} \\ 
\endfoot
\hline
\endlastfoot
Demographical       & 0.482 (0.473, 0.490) & 0.492 (0.483, 0.501) & 0.517 (0.497, 0.537) & 0.467 (0.454, 0.480) & 0.499 (0.488, 0.509) \\ \hline
Biopsy              & 0.607 (0.598, 0.615) & \textbf{0.651} (0.642, 0.661) & \textbf{0.760} (0.739, 0.782) & 0.543 (0.528, 0.557)  & 0.687 (0.677, 0.697) \\ \hline
Manual - all & 0.588 (0.579, 0.596) & 0.553 (0.543, 0.563) & 0.468 (0.447, 0.488) & 0.638 (0.628, 0.649) & 0.575 (0.563, 0.586) \\ \hline
Closing 08 - all & 0.567 (0.558, 0.576) & 0.532 (0.523, 0.542) & 0.446 (0.428, 0.464) & 0.619 (0.607, 0.631) & 0.552 (0.542, 0.563) \\ \hline
Closing 07 - all & 0.591 (0.581, 0.600) & 0.553 (0.543, 0.564) & 0.462 (0.443, 0.481) & 0.645 (0.632, 0.658) & 0.573 (0.562, 0.584) \\ \hline
Closing 06 - all & 0.591 (0.581, 0.601) & 0.555 (0.544, 0.566) & 0.468 (0.448, 0.488) & 0.643 (0.629, 0.656) & 0.578 (0.566, 0.590) \\ \hline
Ellipsoid 04 - all &	0.560 (0.551, 0.569) &	0.516 (0.506, 0.526) & 0.407 (0.390, 0.425)	& 0.625 (0.612, 0.637) &	0.524 (0.514, 0.535)	\\ \hline
Manual - best-SHAP  & 0.643 (0.635, 0.651) & 0.624 (0.615, 0.633) & 0.577 (0.560, 0.595) & 0.671 (0.660, 0.683) & 0.673 (0.664, 0.683) \\ \hline
Closing 08 - best-SHAP & 0.647 (0.638, 0.655) & 0.633 (0.624, 0.642) & 0.598 (0.582, 0.615) & 0.667 (0.655, 0.679) & 0.685 (0.675, 0.695) \\ \hline
Closing 07 - best-SHAP & 0.655 (0.648, 0.663) & 0.638 (0.629, 0.647) & 0.595 (0.575, 0.615) & 0.681 (0.670, 0.692)  & 0.695 (0.686, 0.705) \\ \hline
Closing 06 - best-SHAP & \textbf{0.671} (0.663, 0.678) & 0.649 (0.641, 0.657) & 0.595 (0.577, 0.614) & \textbf{0.703} (0.690, 0.715) & \textbf{0.696} (0.686, 0.706) \\ \hline
Ellipsoid 04 - best SHAP	& 0.639 (0.632, 0.647) &	0.618 (0.610, 0.627) &	0.566 (0.549, 0.583) &	0.671 (0.660, 0.681) &	0.671 (0.662, 0.680) \\ \hline				
\end{longtable}

}

\clearpage

Kolmogorov-Smirnov tests provided evidence against equal model performance for all the comparisons between baseline models and best-SHAP models (for visual reference see Figure \ref{fig:roc_clinical_all_shap}). No statistical significance is observed between the performances of biopsy-based model and best-SHAP models, nor is any difference detected in the pairwise comparison of the best-SHAP models across all masks, except between best-SHAP-closing $07$ and best-SHAP-ellipsoid $04$ models.   
Table \ref{tab:pvalues} provides a summary of these key comparisons. 

  {\fontsize{10}{10}\selectfont
  	 \begin{ThreePartTable}
  	 \begin{longtable}[t]{|llc|}

  		 \caption[Results of the Kolmogorov-Smirnov tests to evaluate differences in model performance. ]{Results of the Kolmogorov-Smirnov tests to evaluate differences in model performance. Bonferroni correction for multiple comparisons was applied. Significance column: '*' indicates statistical evidence against equal model performance after Bonferroni correction; '-' means no statistical significance.}
  		 \label{tab:pvalues} \\
  		 \hline
  		 \textbf{Model 1} & \textbf{Model 2} & \textbf{Significance} \\ \hline
  		 \endfirsthead
  		 \hline
  		 \textbf{Model 1} & \textbf{Model 2} & \textbf{Significance} \\ \hline
  		 \endhead
  		 \hline \multicolumn{3}{r}{\textit{Table \ref{tab:pvalues} continued on next page}} \\ 
  		 \endfoot
  		 \hline
  		 \endlastfoot

		baseline model (any mask) & best-SHAP model (any mask) & * \\ \hline
  		biopsy model & best-SHAP model (any mask) & -\\ \hline
  		best-SHAP model (any mask) & best-SHAP model (any mask) & -\tnote{$\dagger$} \\ \hline
  		\end{longtable}
  		
  		\begin{tablenotes}
  			\item[$\dagger$] Exception: statistical significance between best-SHAP-closing $07$ and best-SHAP-ellipsoid $04$.
  		\end{tablenotes}

		 \end{ThreePartTable}
  	 }

\begin{figure}[h!]
\centering
    \includegraphics[width=0.8\linewidth]{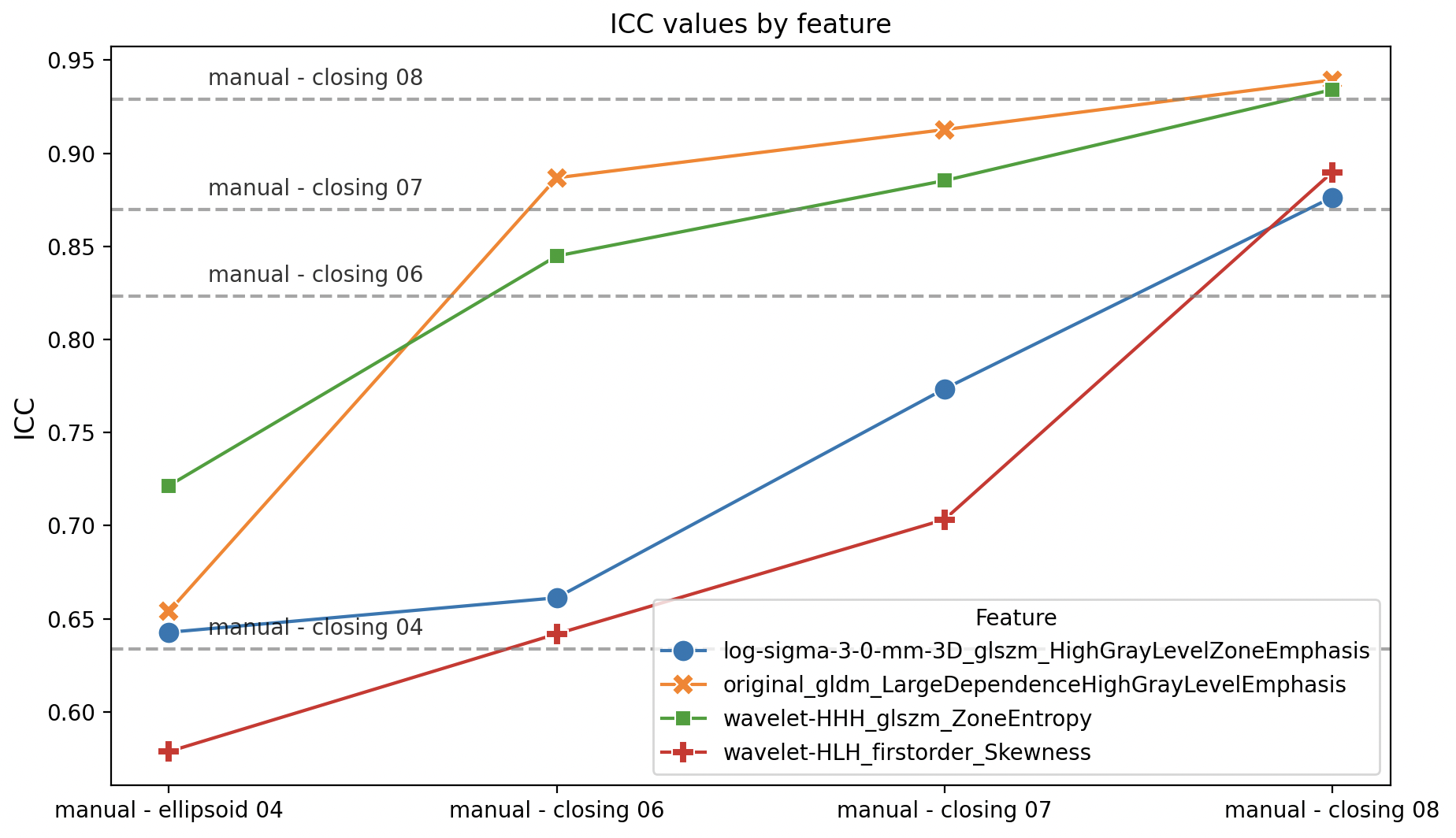}\\
    \includegraphics[width=0.8\linewidth]{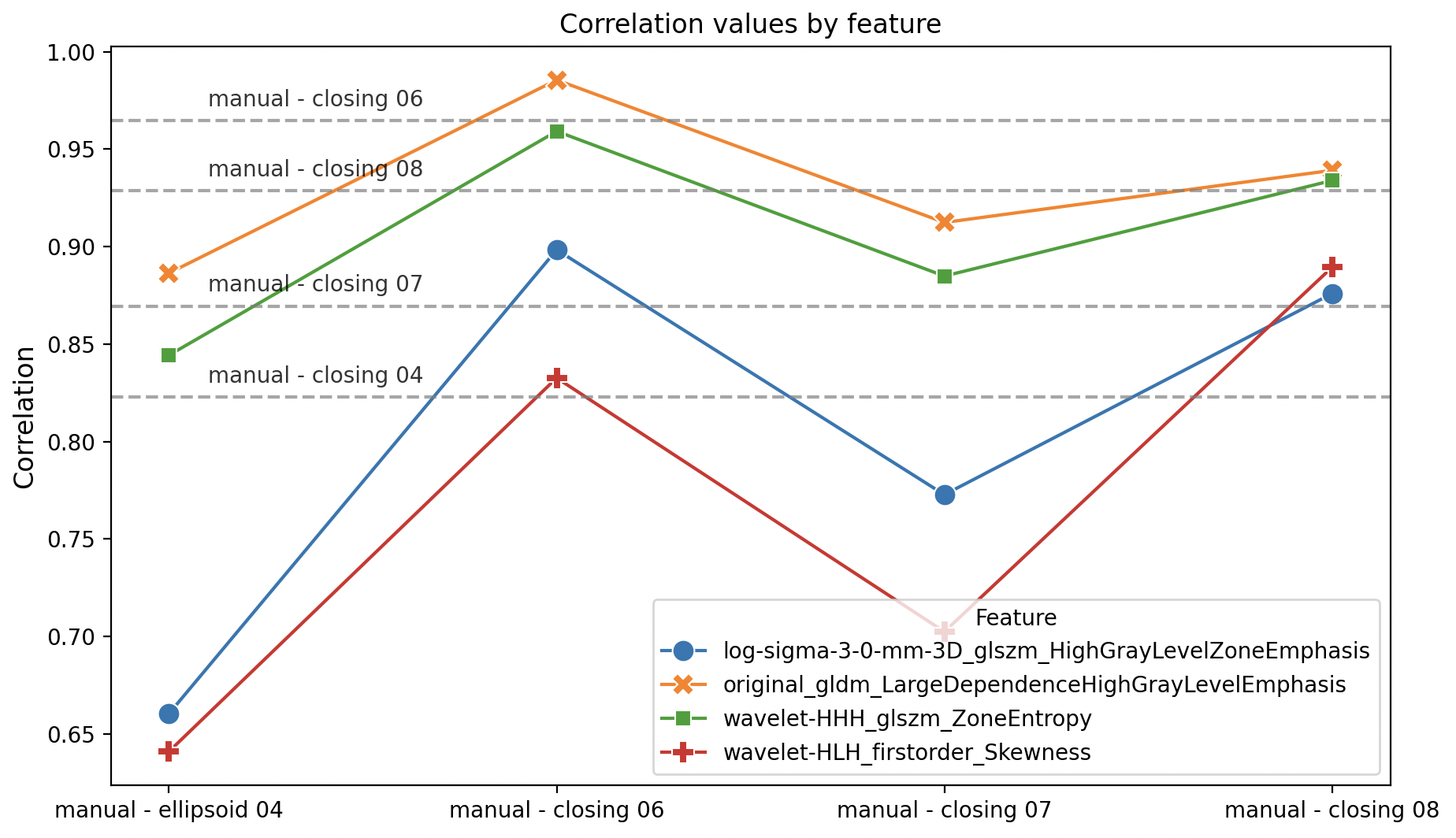}\\
    \caption{Top panel: ICC of the four common best-SHAP features at varying segmentation mask; the dashed lines indicate median ICC on all features, summarizing the reproducibility between features extracted from the manual mask and each of its modifications. Bottom panel: Pearson's correlation of the four common features at varying mask; the dashed lines indicate median correlation on all features, summarizing the correlation between features extracted from the manual mask and each of its modifications.}
    \label{fig:icc_vs_correlation}
\end{figure}

\vspace{0.5cm}
As for the stability analysis, we focused on the four commonly selected best-SHAP features cited above. Figure \ref{fig:icc_vs_correlation} displays the values of ICC and Pearson's correlation for each of these features at varying segmentation mask. The dashed lines indicate the median ICC and Pearson's correlation across all features at varying mask, for comparison with the scores of the four common features.
These features exhibited decreasing ICC values as the segmentation accuracy declined (mean ICC across features for decreasing DSC: 0.910, 0.819, 0.759, 0.649) and high Pearson's correlation coefficient (mean Pearson's correlation across features for decreasing DSC: 0.910, 0.818, 0.919, 0.758).

\begin{figure}[h!]
  	\centering
  	\includegraphics[width=0.8\linewidth]{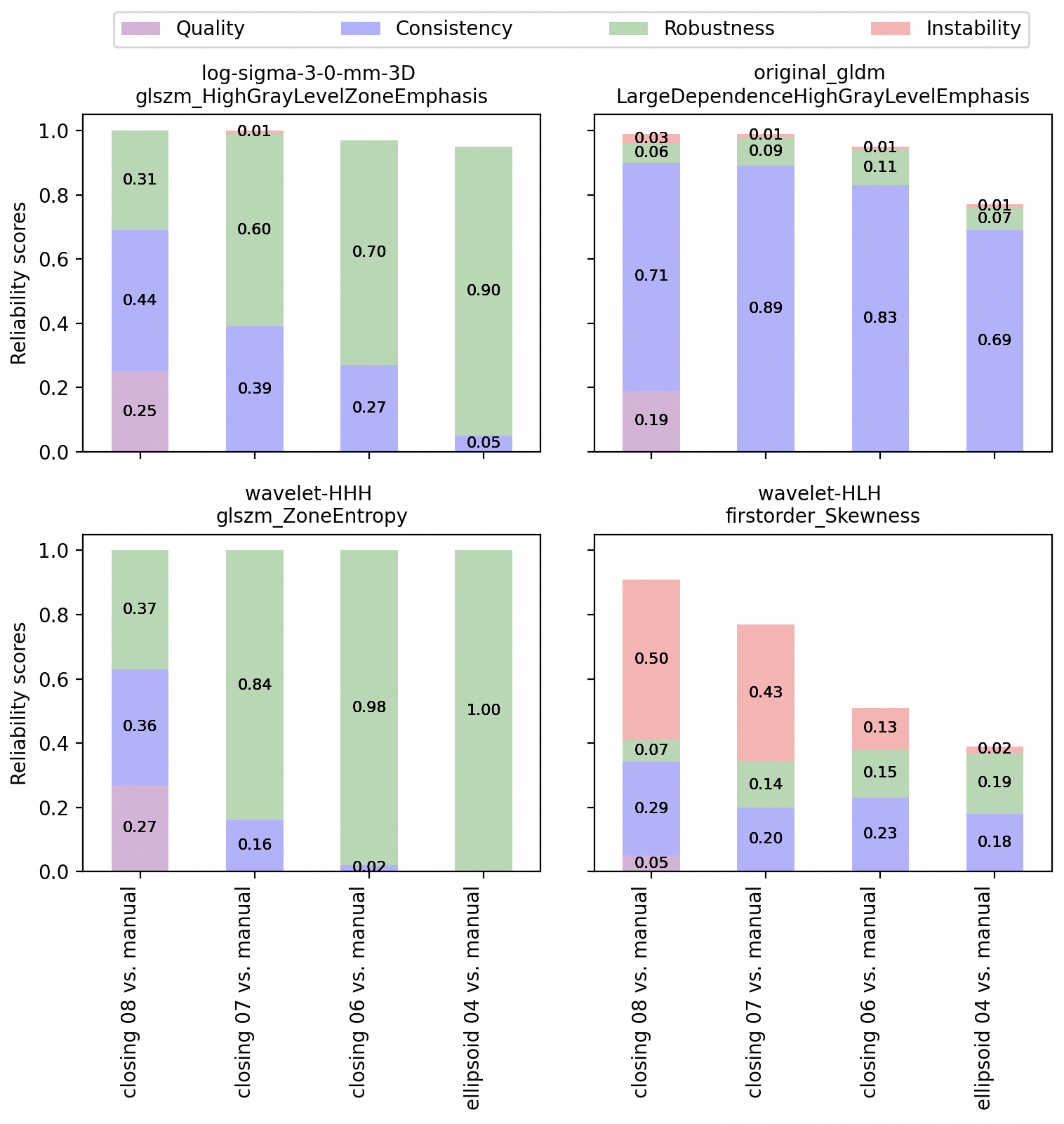}\\
  	\caption[Reliability scores of the four common SHAP features across segmentation masks.]{Reliability scores of the four common SHAP features, computed for each modification of the manual segmentation (manual vs. closing $08$, $07$, $06$, and ellipsoid $04$).}
  	\label{fig:stabilityScoresBest}
  \end{figure}

\begin{figure}[h!]
  	\centering
  	\includegraphics[width=0.9\linewidth]{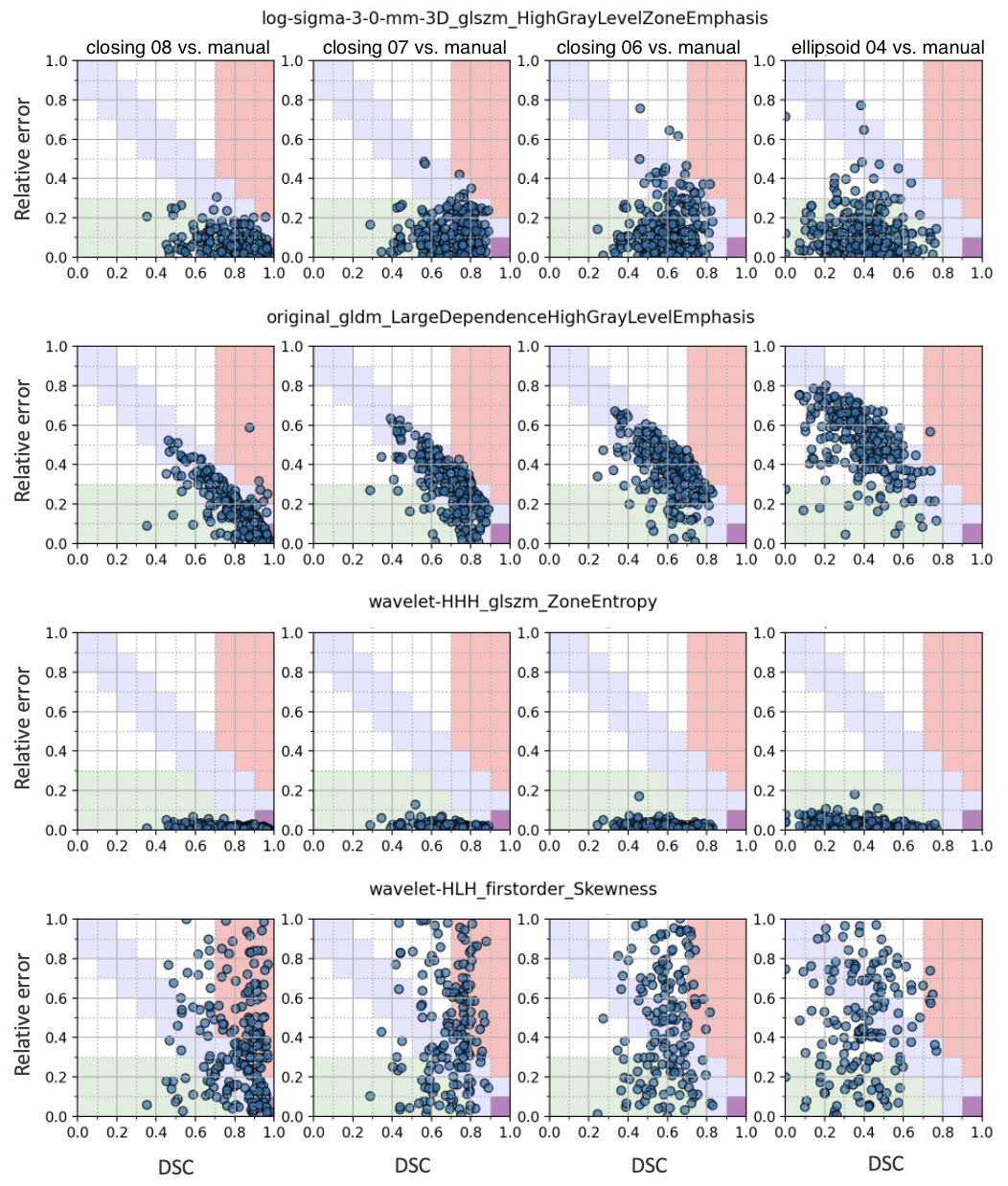}\\
  	\caption[Scatter plots of feature stability w.r.t. segmentation variability across segmentation masks, for the four common SHAP features.]{Scatter plots of relative error on feature value against DSC, for the four common SHAP features, across segmentation masks (from left to right: manual vs. closing $08$, $07$, $06$, and ellipsoid $04$).}
  	\label{fig:matrices_and_scatter}
  \end{figure}

Figure \ref{fig:stabilityScoresBest} reports the reliability/stability scores of the four common SHAP features across mask modification, and Figure \ref{fig:matrices_and_scatter} shows the scatter plots of feature stability w.r.t. segmentation variability for each of these features, for each modification of the manual mask. By definition, the scores may not sum to $1$ if some patients fall outside the predefined ranges of quality, consistency, robustness, and instability. This is particularly evident for the wavelet-HLH-firstorder Skewness feature (Figure \ref{fig:matrices_and_scatter}, last row), as decreasing accuracy of the segmentation mask leads to higher and higher relative error in the feature value for many patients (up to $60$\% of the patients, see bottom right panel of Figure \ref{fig:stabilityScoresBest}), causing them to fall outside the predefined ranges, and highlighting overall feature instability.

\clearpage
\section{Discussion}\label{sec:discussion}
We developed and evaluated ML models for the investigation of the role of radiomic feature in predicting breast cancer subtypes (TNBC vs. non-TNBC) across various segmentation masks.
We used both an extended set of radiomic features generated by a preliminary selection, and a more refined set (best-SHAP features) provided by the SHAP method. Additional models were trained with demographical and biopsy features.

Statistical tests show that no individual feature is able to differentiate TNBC and non-TNBC groups, which remarks the challenge of relying solely on radiomics for the automatic differentiation of breast cancer subtypes.

The results of the ML models show that using a selection of very explicative radiomic features (the best-SHAP features) leads to a prediction performance that is comparable to using biopsy variables, for the binary prediction of tumor subtypes.  Interestingly, prediction results of the best-SHAP models do not exhibit significant statistical difference across segmentation masks, i.e., models trained with features extracted from the manual segmentation and its variations produce comparable performance (see Figure \ref{fig:roc_clinical_all_shap} and Table \ref{tab:pvalues}). 
The prediction obtained for the ellipsoid 04 mask (an ellipsoid over the region of interest provided by the dataset) is particularly surprising. Not only this segmentation mask is extremely poor in terms of DSC, but it also represents a very rough approximation of the tumor shape, missing all contour details and incorporating contrast and irregularity of the surrounding regions of the tumor, like veins, necrosis, edema, and fat tissue (see Figure \ref{fig:overall}, last row).

These results motivated the need to investigate the best-SHAP features selected by each model. We focused our attention on the features that were commonly selected by the five models, as they could elucidate behaviors regarding their stability and robustness across segmentation masks, and offer insights into the factors influencing the prediction results. 
Figure \ref{fig:icc_vs_correlation} compares the ICC and the Pearson's correlation of each common feature, both computed between features extracted from the manual mask and its variations (closing $08$, $07$, $06$, and ellipsoid $04$). As a general trend, the curves associated to original-gldm Large-Dependence High-Gray-Level Emphasis and wavelet-HHH-glszm Zone Entropy (orange and green) are above the ones associated to than LoG $\sigma=3$-glszm High-Gray-Level-Zone Emphasis and wavelet-HLH-firstorder Skewness (blue and red), both for ICC and correlation coefficient. We can notice that, according to thresholds reported in literature \cite{granzier, caballoICC}, LoG $\sigma=3$-glszm High-Gray-Level-Zone Emphasis and wavelet-HLH-firstorder Skewness would have been discarded by the feature selection step (ICC$<0.9$) from the prediction model based on the manual mask. 
However, despite their low ICC values (e.g., lower than 0.75), these features are still considered important and are consistently selected in all predictive models.

This observation suggests that relying solely on ICC for feature selection may not be the most effective approach. Specifically, when examining the correlation coefficient, we find cases where a low ICC does not necessarily correspond to a low correlation coefficient, see Figure \ref{fig:icc_vs_correlation}. This discrepancy arises because the ICC is penalized when the inter-rater variance is high (raters are represented by segmentation masks), while the correlation coefficient ignores inter-rater effects and merely measures the correlation of feature values across different segmentations. 

The high correlation of a predictive feature with its corresponding values extracted from another segmentation mask suggests that both these features are capturing meaningful information relevant to the target, despite the poor reliability rated by ICC.

Additionally, the computation of the reliability scores for the common features, in Figure \ref{fig:stabilityScoresBest}, highlights that there is no shared numerical relationship between feature stability and segmentation variability among the predictive features.
However, segmentation variability does not disrupt feature stability pattern of individual features, as robust, consistent, and unstable patterns remain well-defined across mask modifications, as displayed in Figure \ref{fig:matrices_and_scatter}.

These considerations suggest that the feature selection process in predictive models is not inherently linked to feature stability with respect to segmentation variability, neither from a numerical nor from a reproducibility perspective (ICC), but rather to general patterns captured by the distribution of features across patients. 
This underscores why feature selection approaches based on ICC or reliability scores may fail to identify the most predictive features for a specific task.
On the other hand, the high correlation between the same predictive features extracted from different segmentation masks helps explaining why the best-SHAP models yield similar performance.

Overall, the results here presented suggest that high segmentation accuracy may not be an imprescindible requirement for radiomics applications, and that including partial peritumoral imaging information through various segmentation masks does not hamper feature predictive power. \\

This study presents some limitations. First, the variability range of the ROC-AUC is not particularly narrow, a concern previously highlighted by Montoya-del-Angel et al. \cite{marti} in other radiomics-based applications for breast cancer. Second, the dataset we used was obtained from a single institution, therefore no independent testing was conducted. Although cross-validation helps reducing overfitting on specific test sets, future work should involve validating the method with larger, multi-institutional datasets to better assess reproducibility and generalizability of the results.
Additionally, the set of extracted first-post-contrast texture features and tumor morphology features may not adequately capture the full spectrum of complexities involved in tumor subtype classification. Future investigations should incorporate time-dependent texture descriptors and heterogeneity features related to enhancement kinetics into the ML models, as illustrated in Caballo et al. \cite{caballo2023four} for other prediction purposes.
Future work may also include features extracted from other MRI sequences, e.g., Diffusion Weighted Imaging (DWI) and ADC, to further improve cancer characterization.

\section{Conclusions}\label{sec:conclusions}

This study explored the performance of radiomics-based ML models designed to distinguish TNBC from other tumor subtypes across variable segmentation masks, achieved through various modifications of the manual segmentation of breast cancer lesions.  
According to the findings reported in this paper, the prediction of TNBC subtype results to be independent across variable breast cancer segmentation masks. This suggests that achieving precise segmentation accuracy may not be a necessary prerequisite for radiomic applications and that incorporating partial peritumoral information does not necessarily compromise the feature predictive capability. 
Moreover, results on reproducibility metric and reliability scores computed for the most predictive features suggest that applying feature selection techniques based on reproducibility across segmentations, or aimed at rewarding feature stability w.r.t. segmentation accuracy, may lead to the exclusion of predictive features from the model input. 
It is possible that, if radiomics exhibited greater predictive power, the impact of segmentation variability on prediction performance would be more evident. However, radiomics predictive capability remains modest, partly due to limited data but mainly because predictive models struggle to consistently identify universally predictive features across different data splits of the same dataset without overfitting. This limitation underscores the importance of using methods for collective explainable feature selection. 

With these findings, this work aims to contribute to a deeper understanding of the intricate interplay between feature stability in relation to segmentation and prediction accuracy, fostering the development of more reliable, generalizable, and clinically applicable image-driven models for breast cancer diagnosis and prognosis.

\subsection*{Disclosures}
All authors declare no conflicts of interest in this paper.

\subsection* {Acknowledgments}
This project has received funding from the European Union’s Horizon Europe and Horizon 2020 research and innovation program under grant agreement No 101057699 (RadioVal) and No 952103 (EuCanImage), respectively. Also, this work was partially supported by the project FUTURE-ES (PID2021-126724OB-I00) and AIMED (PID2023-146786OB-I00) from the Ministry of Science, Innovation and Universities of the Government of Spain. CC acknowledges the support of the PRIN PNRR 2022 Project 'Computational mEthods for Medical Imaging (CEMI)' 2022FHCNY3, cup: D53D23005830006. MP acknowledges the financial support of the "Hub Life Science – Digital Health (LSH-DH) PNC-E3-2022-23683267 - Progetto DHEAL-COM – CUP: D33C22001980001". Also, this research was supported in part by the MIUR Excellence Department Project awarded to Dipartimento di Matematica, Università di Genova, CUP D33C23001110001. IC, CC, MP, and SG are members of the Gruppo Nazionale per il Calcolo Scientifico - Istituto Nazionale di Alta Matematica (GNCS - INdAM). The work is partially supported by INdAM - GNCS Project “Analisi e applicazioni di matrici strutturate (a blocchi)", CUP E53C23001670001.

\subsection* {Data, Materials, and Code Availability} 
Data used for the analysis can be downloaded from Synapse at https://doi.org/10.7303/syn60868042 as part of the Duke dataset \cite{duke_dataset, Garrucho2024MAMAMIA}. The exact list of images and the codes used for the analysis are available upon reasonable request.

\bibliography{references}
\bibliographystyle{spiejour}

\listoffigures
\listoftables

\end{document}